\documentclass[letterpaper, 10 pt, conference]{ieeeconf}  

\IEEEoverridecommandlockouts                              

\usepackage{url}
\usepackage{graphicx}

\usepackage{epstopdf}
\usepackage[font=footnotesize]{caption}
\usepackage[font=footnotesize]{subcaption}

\usepackage{color}
\usepackage{transparent}

\usepackage{multirow}

\title{\LARGE \bf
The Shape of Health: A Comparison of Five Alternative Ways of Visualizing Personal Health and Wellbeing*.
}

\author{Andres Ledesma$^{1}$, \textit{Student Member IEEE}, Hannu Nieminen$^{1}$, P\"{a}ivi Valve$^{1, 2}$,\\
Miikka Ermes$^{3}$, Holly Jimison$^{4}$, \textit{Member IEEE} and  Misha Pavel$^{4}$, \textit{Senior Member IEEE}
    \thanks{*This research was supported jointly by TEKES (the Finnish Funding Agency for Technology and Innovation) as part of the Digital Health Revolution and FiDiPro (Finland Distinguished Professor Programme) projects and the European Commission and TEKES under the ARTEMIS-JU WithMe project.}
    \thanks{$^{1}$Andres Ledesma, P\"{a}ivi Valve and Hannu Nieminen are with the Personal Health Informatics research group as part of the Department of Signal Processing, Tampere University of Technology, Tampere, Finland}%
  		\thanks{{\tt\small andres.ledesma@tut.fi,}}
  		\thanks{{\tt\small hannu.o.nieminen@tut.fi, paivi.valve@tut.fi}}%
    \thanks{$^{2}$P\"{a}ivi Valve is also with the School of Health Sciences, University of Tampere,
          Tampere, Finland}%
    \thanks{$^{3}$Miikka Ermes is with VTT Technical Research Centre of Finland, Tampere, Finland}
		\thanks{{\tt\small miikka.ermes@vtt.fi}}%
    \thanks{$^{4}$Misha Pavel and Holly Jimison are with the College of Computer \& Information Science and the Bouv\'{e} College of Health Sciences, Northeastern University, Boston MA, USA}
		\thanks{{\tt\small m.pavel@neu.edu, h.jimison@neu.edu}}%
}

\begin{document}

 \maketitle
\thispagestyle{empty}
\pagestyle{empty}

\begin{abstract}

The combination of clinical and personal health and wellbeing data can tell us much about our behaviors, risks and overall status. The way this data is visualized may affect our understanding of our own health. To study this effect, we conducted a small experiment with 30 participants in which we presented a holistic overview of the health and wellbeing of two modeled individuals, one of them with metabolic syndrome. We used an insight-based methodology to assess the effectiveness of the visualizations. The results show that adequate visualization of holistic health data helps users without medical background to better understand the overall health situation and possible health risks related to lifestyles. Furthermore, we found that the application of insight-based methodology in the health and wellbeing domain remains unexplored and additional research and methodology development are needed.

\end{abstract}

\section{INTRODUCTION}

Modern medicine is constantly moving towards a preventive approach. As stated by Rose \cite{rose1992strategy}, ``common diseases have their roots in life-style, social factors and the environment, and successful health promotion depends upon a population-based strategy of prevention''. Self-management of health and wellness plays a key role in making this happen. 

A holistic description of health should ideally combine together several measures of physical, mental and social wellbeing \cite{who1948health}. Individuals already start to have access to a wealth of multimodal health-related data e.g. from personal wellness devices, health records, genetic tests and mobile wellness applications. This information, when combined together, can be utilized both to empower people to monitor their evolving health \cite{shneiderman2013improving} and to enhance the healthcare process. However, understanding the status of a person's health and the underlying factors behind the data is not easy, especially for the non-experts. Presenting this information in a comprehensive and meaningful way is an ongoing effort and an open challenge \cite{shneiderman2013improving}. 

As an example, TimeLine \cite{bui2007information} is a software tool that organizes medical records and provides a ``problem-centric temporal visualization''. Lesselroth and Pieczkiewicz \cite{lesselroth2011data} conducted an extensive literature survey on strategies for the visualization of personal health data. They argue that ``smart dashboards'' combining different data sources are needed to improve the understanding of our health. Although various graphs, stylized presentation and textual feedback have been addressed in studies \cite{Consolvo2008ASW}, visualizations combining multiple data sources that enable inferences are mostly lacking \cite{shneiderman2013improving}. Health applications enabling to identify connections that are significant over time have been shown to increase the user's self-understanding \cite{bentley2013health}. Goetze \cite{1_goetz_2015} has shown that the style of the presentation of data greatly affects the understanding of our health. The hGraph \cite{follett2012hgraph} is an example of a visual representation of a patient's health status, designed to increase awareness of the factors that can affect one's overall health. 



Graphical perception, defined as ``the visual decoding of information encoded on graphs'' \cite{cleveland1984graphical}, has been widely studied and researched \cite{baird1978fundamentals}. This paper focuses on evaluating how the participants identify the underlying reasons of the health situation of a person represented as a set of measurements. We study how the visual representation of holistic health data affects the capability of non-medical experts to derive insights. Insights aim to see ``beyond the figures'' by identifying relationships between the measurements and understanding the underlying reasons for the values. We also consider the effect of health literacy in this study, which is defined as ``the degree to which individuals have the capacity to obtain, process, and understand basic health information and services needed to make appropriate health decisions'' \cite{baur2010national}.


\begin{figure*}[th]
\centering
\par\bigskip\bigskip
\begin{subfigure}{.2\linewidth}
\centering
\def\svgwidth{.7\linewidth}
{\renewcommand\normalsize{\scriptsize}%
\normalsize
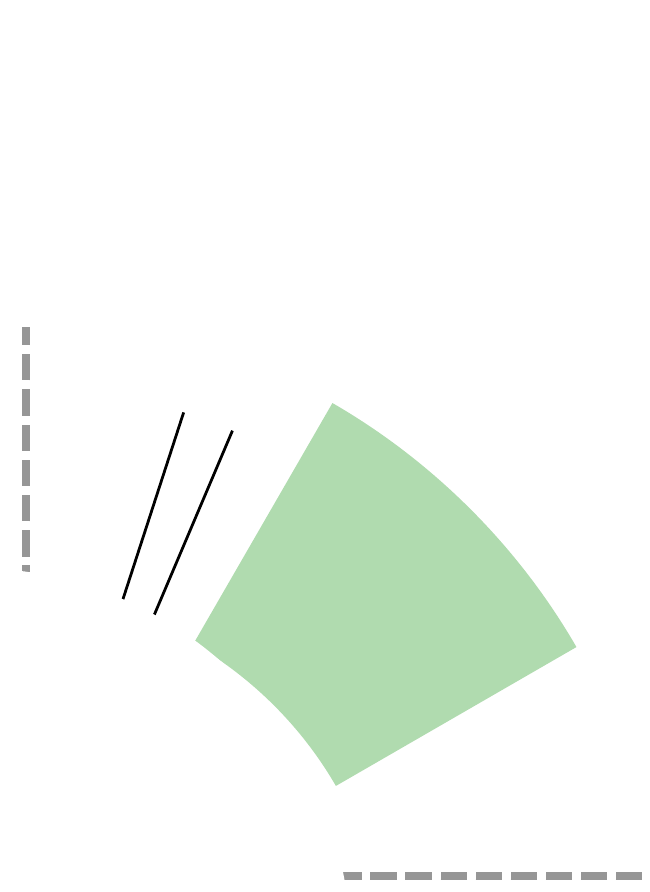}%
\subcaption{Angle}
\label{fig:angle}
\end{subfigure}
\hspace{.1cm}
\begin{subfigure}{.175\linewidth}
\centering
\def\svgwidth{.8\linewidth}
{\renewcommand\normalsize{\scriptsize}%
\normalsize
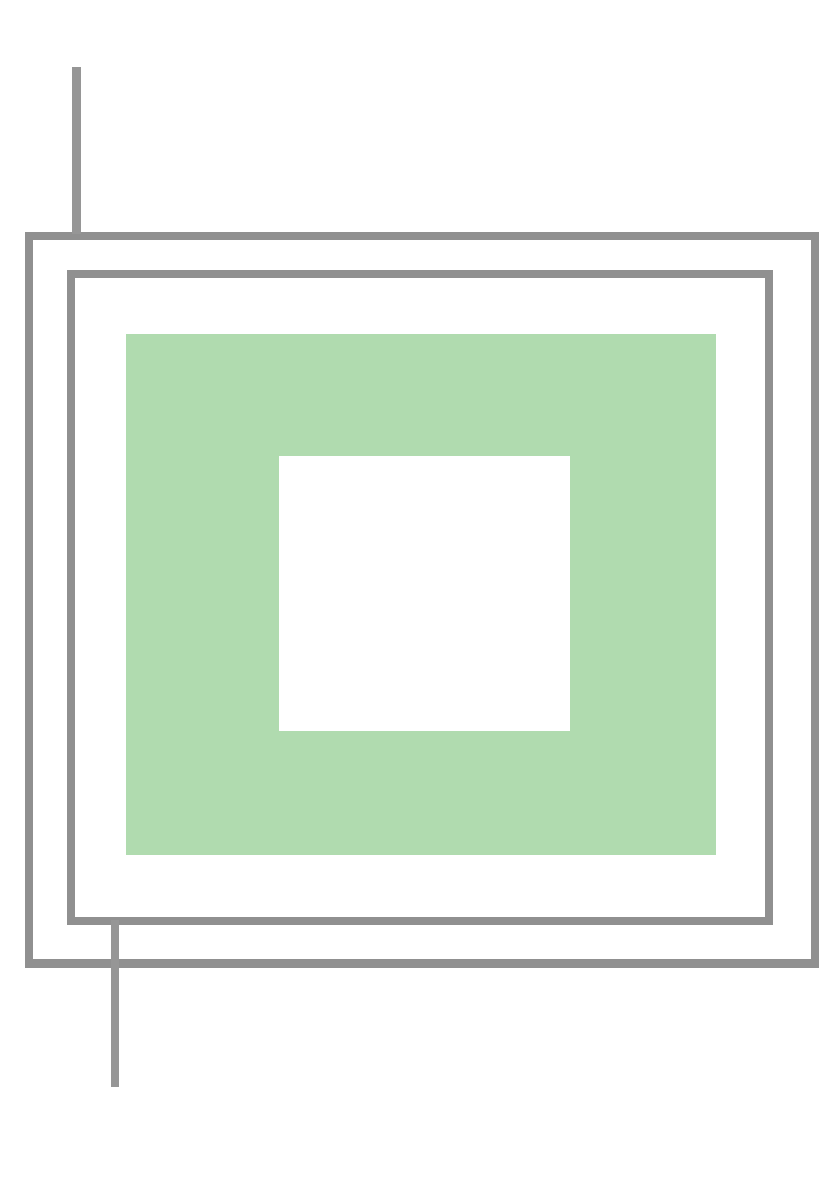}%
\subcaption{Area}
\label{fig:area}
\end{subfigure}
\hspace{.2cm}
\begin{subfigure}{.275\linewidth}
\centering
\def\svgwidth{.95\linewidth}
{\renewcommand\normalsize{\scriptsize}%
\normalsize
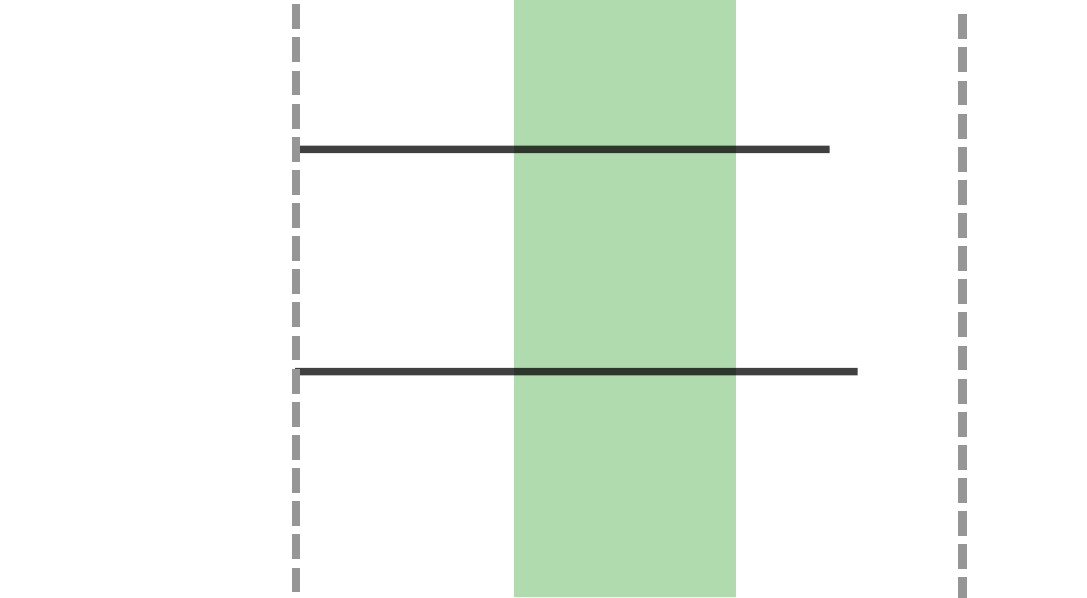}%
\subcaption{Length}
\label{fig:length}
\end{subfigure}
\hspace{.1cm}
\begin{subfigure}{.275\linewidth}
\centering
\def\svgwidth{.95\linewidth}
{\renewcommand\normalsize{\scriptsize}%
\normalsize
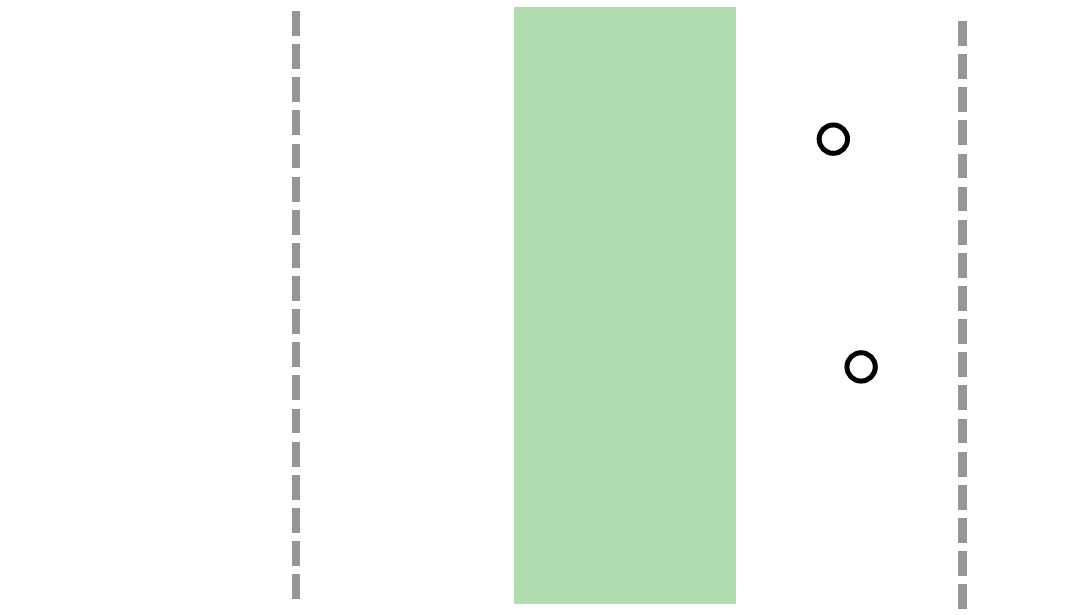}%
\subcaption{Position}
\label{fig:position}
\end{subfigure}
\caption{The blood pressure measurements represented using the angle, area, length and position.}
\label{fig:visualizations}
\vspace{-1.5em}
\end{figure*}

\section{METHODS}

We recruited 15 female and 15 male participants from 9 different countries. We used the local bulletin board to advertise the study. Participants having earlier expertise in health data analysis and visualization were excluded from the study. The students signed up using a Doodle poll. The experimental procedures described in this paper complied with the principles of Helsinki Declaration of 1975, as revised in 2000. All subjects gave informed consent to participate and they had a right to withdraw from the study at any time. Their information was anonymized prior the analysis. The participants received a movie ticket at the end of the session.

	\subsection{Datasets Used in the Study}

We modeled two datasets of measurements to describe two different health situations. The first dataset modeled a person with a poor overall health condition with a clear indication of a metabolic syndrome and therefore a high to very high risk of developing Type 2 Diabetes (T2D), based on the calculator in \cite{lindstrom2003diabetes}. The second set modeled a person with a good health condition and with a healthy lifestyle. The participants focused on the modeled person with metabolic syndrome to derive their insights, and the other person was used as a comparison point.


The following health parameters (average values of past month) were chosen to describe their health and wellbeing:
\begin{itemize}
	\item Blood pressure: systolic and diastolic blood pressure 
    \item Physical activity: weekly active days, steps per day 
    \item Body composition: Body Mass Index (BMI), waist diameter and fat percentage
    \item Sleep: time in bed, time asleep  
    \item Fitness: resting heart rate, fitness index, muscular force, muscular endurance and balance \cite{laukkanen1992validity} 
    \item Lab Tests: hemoglobin, fB-Gluc, cholesterol, HDL, LDL, triglycerides 
    \item Drugs: tobacco (cigarettes per day), alcohol abuse, drug abuse (narcotics), medication abuse
    \item Emotional wellbeing: depression level (DEPS), stress level and stress recovery \cite{firstbeat2014} as well as optimism \cite{scheier1994distinguishing}.
\end{itemize}

Healthy ranges for each parameter, derived based on the national clinical health recommendations \cite{Kaypahoito}, were visualized with a light green background. This concept was also explained to the study participants. 

\subsection{Visualizations}


We utilized the generic visualization framework proposed by Cleveland and McGill \cite{cleveland1984graphical} to design four different visualizations (Fig. \ref{fig:visualizations}). In addition, we included the polar coordinate style ``hGraph'' \cite{follett2012hgraph} visualization, to assess it against the framework of Cleveland and McGill. 

\begin{figure}[th]
\centering
\def\svgwidth{.8\linewidth}
{\renewcommand\normalsize{\footnotesize}%
\normalsize
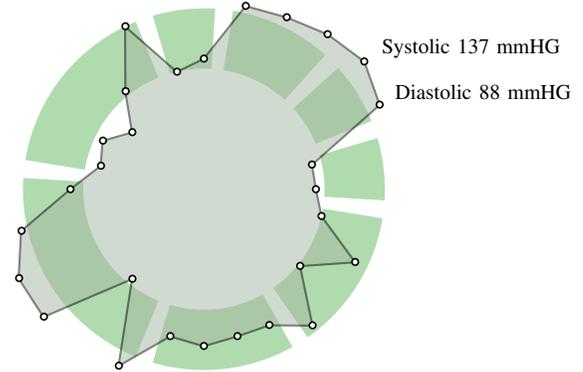}%
\caption{The hGraph representation of the data with two example labels.}
\label{fig:hgraph}
\vspace{-1.5em}
\end{figure}

		\subsubsection{Cleveland and McGill Framework}

\paragraph{Angle} The degree of the angle represents the value of the measurements. Figure \ref{fig:angle} shows the lines with an angle as a representation of the blood pressure.

\paragraph{Area} The size of a shape represents the value of the measurements. Figure \ref{fig:area} shows the blood pressure represented as squares where the size represents the value.

\paragraph{Length} The length of the line represents the value of the measurements. Figure \ref{fig:length} shows the length of the lines representing the values of the blood pressure.

\paragraph{Position Along Aligned Scales} The value of a measurement is represented by the position. The position of the circles in figure \ref{fig:position} represent the values of the blood pressure.

		\subsubsection{hGraph}
        
The design emphasizes the presentation of a large number of parameters and how they conform a holistic overview of the health of a person. Figure \ref{fig:hgraph} shows the hGraph of the modeled data with metabolic syndrome. Multiple parameters are arranged as polar coordinates and a figure connecting the coordinates shows how well the measurements are with respect to the recommended values (green area).

		\subsubsection{Control Group}
        
The control group received a table with numeric values. These participants relied only on numeric values and recommended ranges.

\begin{figure*}[tbh]
\centering
\par\bigskip
\begin{subfigure}{.15\linewidth}
\centering
\def\svgwidth{.8\linewidth}
{\renewcommand\normalsize{\scriptsize}%
\normalsize
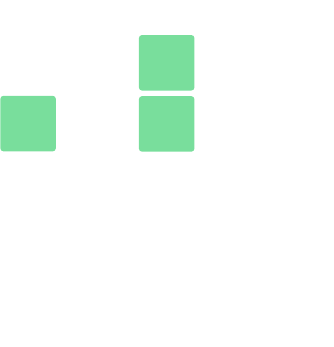}%
\subcaption{Angle}
\label{fig:assessment-angle}
\end{subfigure} 
\begin{subfigure}{.15\linewidth}
\centering
\def\svgwidth{.8\linewidth}
{\renewcommand\normalsize{\scriptsize}%
\normalsize
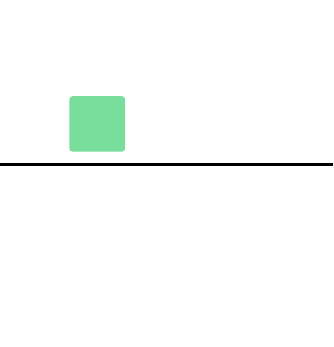}%
\subcaption{Area}
\label{fig:assessment-area}
\end{subfigure} 
\begin{subfigure}{.15\linewidth}
\centering
\def\svgwidth{.8\linewidth}
{\renewcommand\normalsize{\scriptsize}%
\normalsize
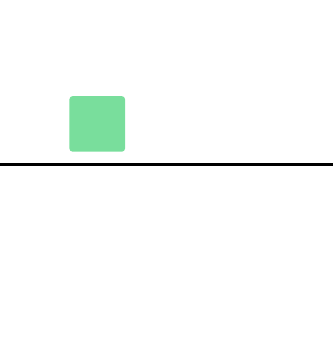}%
\subcaption{hGraph}
\label{fig:assessment-hGraph}
\end{subfigure} 
\begin{subfigure}{.15\linewidth}
\centering
\def\svgwidth{.8\linewidth}
{\renewcommand\normalsize{\scriptsize}%
\normalsize
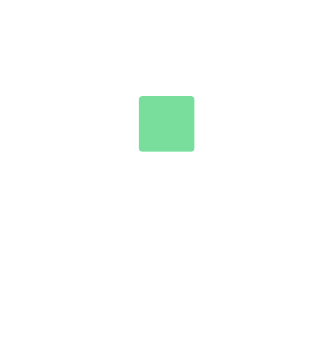}%
\subcaption{Length}
\label{fig:assessment-length}
\end{subfigure} 
\begin{subfigure}{.15\linewidth}
\centering
\def\svgwidth{.8\linewidth}
{\renewcommand\normalsize{\scriptsize}%
\normalsize
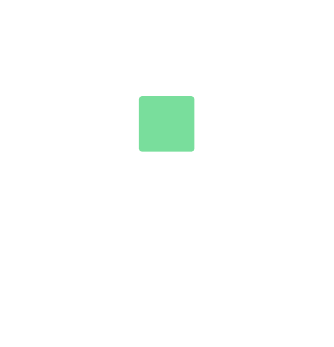}%
\subcaption{Position}
\label{fig:assessment-position}
\end{subfigure} 
\begin{subfigure}{.15\linewidth}
\centering
\def\svgwidth{.8\linewidth}
{\renewcommand\normalsize{\scriptsize}%
\normalsize
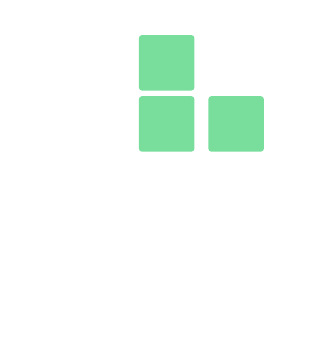}%
\subcaption{Table}
\label{fig:assessment-table}
\end{subfigure} 
\caption{Assessments of overall health per visualization, ordered left from right according to the subjects'  Health Literacy, i.e. familiarity on the risks of T2D (from ``Very familiar'' to ``Not familiar at all''). Green refers to a correct and red to an incorrect assessment.}
\label{fig:assessment}
\vspace{-1.5em}
\end{figure*}

	\subsection{Experiment Protocol}


The experiment started with a self-assessment of participants' knowledge on health and wellbeing, i.e. their familiarity with health and risk behaviors (diet, sleep, physical activity, etc) and how they affect one's health and wellbeing. Additionally, we asked the participants how familiar they think they are with the risks of developing T2D.


Block randomization, where only one of the visualizations (or the table) was shown to each subject, was utilized. We used a web browser to render the graphics (JavaScript and SVG). The whole session was recorded with a video camera.

We applied the insight-based methodology proposed by North \cite{north2006toward}. We asked the participants to ``see beyond the figures'' and explain the relationships between the measurements in order to gain understanding of the health situation as well as to identify the possible underlying reasons. We asked the subject to tell us as many insights as possible within ten minutes. Participants were encouraged to use a think aloud process. 

%


Afterwards, we asked the participants to perform an investigative analysis \cite{Youn2009Evaluating} by assessing the overall health and wellbeing as well as the  risks of developing T2D of the two datasets. We used a scale of five for both tasks. The options for the overall health and wellbeing assessment ranged from  \textit{``Very Poor Health''} to \textit{``Very Good Health''} and for the T2D risk from \textit{``Very High Risk''} to \textit{``Very Low Risk''}.

\begin{table}[b]\small
\centering
\footnotesize
\vspace{-1.75em}
\caption{Proportion of correct assessments of the T2D risk and overall health at different levels of self-assessed familiarity with T2D risk factors (Health Literacy).}
\begin{tabular}{l c c}
\hline
Familiarity & Risk (\%) & Health (\%) \\ \hline
Very familiar (1) & 100\% & 100\% \\ \hline
Familiar (8) & 100\% & 75\% \\ \hline
Somewhat familiar (10) & 90\% & 80\% \\ \hline
Slightly familiar (8) & 75\% & 50\% \\ \hline
Not familiar at all (3) & 66\% & 33\% \\ \hline
\end{tabular}
\label{table:correctAssessments}
\end{table}





We used depth and complexity \cite{north2006toward} to evaluate the insights. Similar to \cite{saraiya2004evaluation}, we used a five-point scale. Clinically incorrect insights did not give any points. Superficial insights that stated e.g. that a value is low or high were rated as one point. Assumptions and relationships between measurements were rated as a value of two. For instance, we assigned two points if the participant mentioned that some measurements are related or that ``the person does not exercise'' with no explanation. Explanations of the insights were rated as three points. The explanations had complexity, depth or both. For our case, complexity refers to the relationships between measurements and depth to the underlying reasons of the values. When a participant formulated a hypothesis that described the relationship of measurements and the possible reasons behind the values, we evaluated the insight with four points. In order to obtain five points, the participants had to relate all the measurement groups in a single hypothesis explaining the underlying reasons for the values. The insights were evaluated blindly by the authors of this article following the before-mentioned criteria. 


\section{RESULTS AND DISCUSSION}

Correct assessment of T2D risk and overall health depended on the participants' familiarity with T2D risk factors (see table \ref{table:correctAssessments}). The correct assessment of overall health and the general knowledge on health and wellbeing did not have such dependency (data not shown). Therefore we selected familiarity with T2D risk factors as our measure of Health Literacy.



Figure \ref{fig:assessment} shows the assessments of the overall health with different visualization techniques. The \textit{area} had the least correct assessments as shown in figure \ref{fig:assessment-area}. Figure \ref{fig:assessment-angle} and \ref{fig:assessment-hGraph} show that the \textit{angle} and \textit{hGraph} supported participants with low Health Literacy to assess the dataset correctly, as compared with \textit{lenght} and \textit{position}. Additionally, figure \ref{fig:assessment-table} shows that the \textit{table} also supported correct assessments.

\begin{figure}[h]
\centering
\begin{subfigure}{.98\linewidth}
\centering
\vspace{-0.75em}
\subcaption{Angle}
\def\svgwidth{.98\linewidth}
{\renewcommand\normalsize{\footnotesize}%
\normalsize
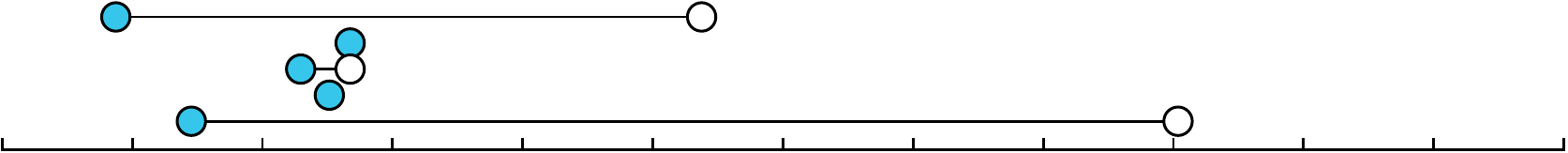}%
\label{fig:timeAngle}
\end{subfigure} 
\par\medskip
\begin{subfigure}{.98\linewidth}
\centering
\subcaption{Area}
\def\svgwidth{.98\linewidth}
{\renewcommand\normalsize{\footnotesize}%
\normalsize
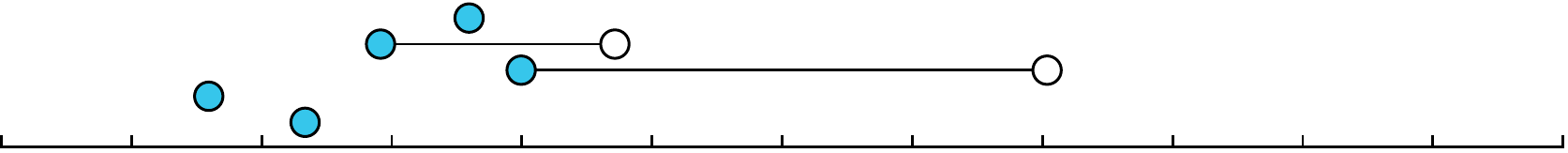}%
\label{fig:timeArea}
\end{subfigure} 
\par\medskip
\begin{subfigure}{.98\linewidth}
\centering
\subcaption{hGraph}
\def\svgwidth{.98\linewidth}
{\renewcommand\normalsize{\footnotesize}%
\normalsize
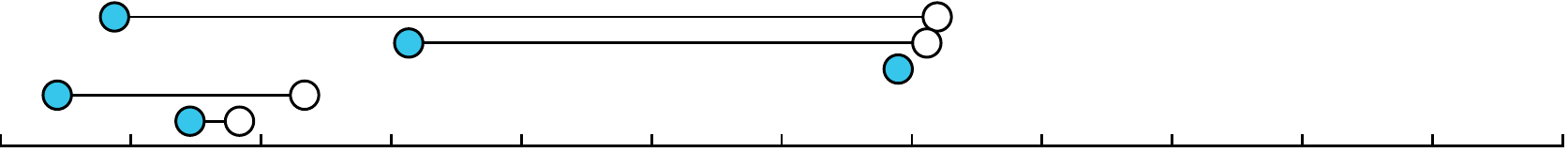}%
\label{fig:timehGraph}
\end{subfigure} 
\par\medskip
\begin{subfigure}{.98\linewidth}
\centering
\subcaption{Length}
\def\svgwidth{.98\linewidth}
{\renewcommand\normalsize{\footnotesize}%
\normalsize
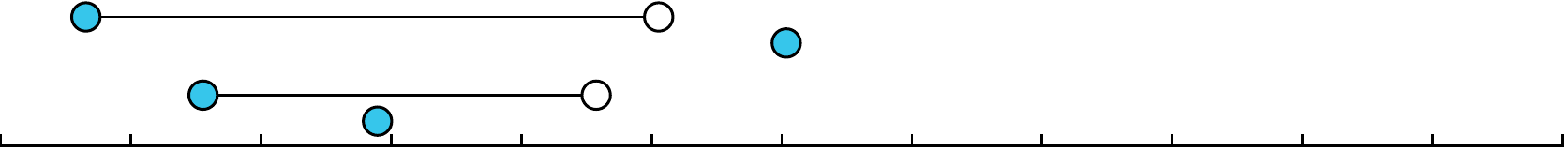}%
\label{fig:timeLength}
\end{subfigure} 
\par\medskip
\begin{subfigure}{.98\linewidth}
\centering
\subcaption{Position}
\def\svgwidth{.98\linewidth}
{\renewcommand\normalsize{\footnotesize}%
\normalsize
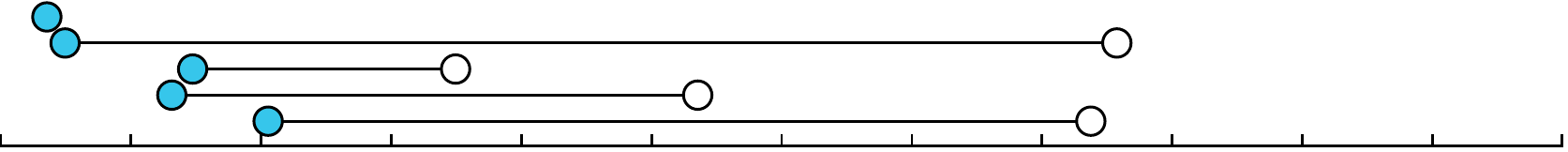}%
\label{fig:timePosition}
\end{subfigure} 
\par\medskip
\begin{subfigure}{.98\linewidth}
\centering
\subcaption{Table}
\def\svgwidth{.98\linewidth}
{\renewcommand\normalsize{\footnotesize}%
\normalsize
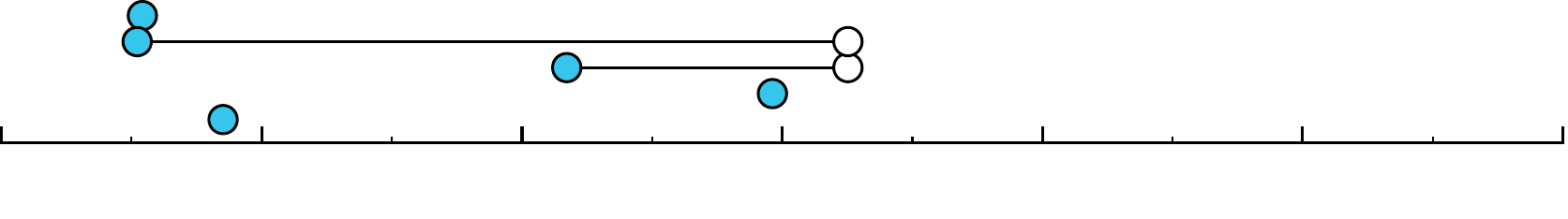}%
\label{fig:timeTable}
\end{subfigure} 
\caption{Time to first insight value three and time to first hypothesis are shown as blue and white circles respectively.}
\label{fig:time}
\vspace{-.65em}
\end{figure}

As in Saraiya et al. \cite{saraiya2004evaluation}, we computed the time to the first insight. We used the time to first insight of value three, namely when a participant was able to give explanations substantiating the observation. Figure \ref{fig:time} shows the time to first insight of value three as a dot and the time to the first generated hypothesis (insight values as 4 or 5) as a circle. In some cases participants were not able to generate hypothesis or formulate insights of value three at all (as in figure \ref{fig:timeLength}). 



Four out of five participants that used the \textit{position} and \textit{hGraph} generated hypothesis, figure \ref{fig:timehGraph} and \ref{fig:timePosition}. Participants using the \textit{hGraph} generated six hypotheses in total, while participants with the \textit{angle} and \textit{position} generated a four each, see table \ref{table:overall}. The participants using the \textit{table} generated a total of four hypotheses, however one participant generated 3 of them.

The average time to insights value three or more is also shown on table \ref{table:overall}. The longest average time occurred using the \textit{area}, \textit{table} and \textit{length}. The speed is similar to Cleveland and McGill's results \cite{cleveland1984graphical} except for the \textit{length}. However, the lowest Health Literacy was observed in this group.

\begin{table}[h]\small
\centering
\footnotesize
\vspace{-0.75em}
\caption{The table shows the average time to first insight of value three or more, the average Health Literacy of the participants, the number of hypothesis (insights value four or five) in total and per participant.}
\begin{tabular}{c c p{1.1cm} c c}
\hline
\multirow{2}{*}{Visualization }& \multirow{2}{1.1cm}{Time (s)}& \multirow{2}{1.75cm}{Health Literacy}& \multicolumn{2}{ c }{Hypothesis} \\  & & & Total & Per participant \\ \hline
Angle    &  98 & 3.2 & 4 & (0,0,2,1,1) \\ \hline
Area     & 144 & 3.0 & 3 & (0,0,2,1,0) \\ \hline
hGraph   & 128 & 3.2 & 6 & (2,0,1,2,1) \\ \hline
Length   & 139 & 2.0 & 3 & (2,0,1,0,0) \\ \hline
Position &  57 & 2.6 & 4 & (0,1,1,1,1) \\ \hline
Table    & 141 & 3.2 & 4 & (0,0,1,0,3) \\ \hline
\end{tabular}
\label{table:overall}
\vspace{-1.15em}
\end{table}

The evaluation methodology gives low points to obvious insights, such as ``physical activity seems to be low, the patient should start exercising more''. However, such insight is clinically valuable. Therefore, future work is needed to further develop the methodology.

\section{LIMITATIONS}



All the participants in our experiment had an academic degree and in some cases an engineering background. The sample of 30 participants cannot be representative of a wider population.


\section{CONCLUSIONS AND FUTURE WORK}





In this paper we applied the insight-based methodology for the health data visualizations. Adequate visualization of holistic health data was shown to help users without medical background to better understand the overall health situation and possible health risks related to lifestyles. Visualizations could also be valuable tools for the medical experts to enhance the healthcare process.








At the end of the experiment session, participants selected three visualizations that would have helped them better perform in the previous tasks. We also administered a short form of Raven Advanced Progressive Matrices test \cite{arthur1994development}. Further study will be addressed in future research.




We evaluated the insights using an informed but subjective assessment based on the principles from North \cite{north2006toward}. For future work, we aim to measure the insight value by modeling prior and derived knowledge. This model may provide a more accurate and reliable estimate of the performance of different health data visualizations. The goal is to provide customizable visualizations to support better understanding of personal health and wellbeing.








\bibliographystyle{IEEEtran}
\bibliography{IEEEabrv,mybibfile}

\end{document}